\documentclass[         %
aps,                    
prd,                    
showpacs,               
superscriptaddress,    
nofootinbib,            
onecolumn,             %
showkeys,               %
preprintnumbers,        %
floatfix               
]{revtex4}               
\usepackage{graphicx,amsmath}
\usepackage[justification=RaggedRight]{caption}
\usepackage{enumerate}
\usepackage{xcolor}
\usepackage{float}
\usepackage[section]{placeins}
\usepackage{subfigure}

\begin{document}
\title{Matter Neutrino Resonance Transitions above a 
Neutron Star Merger Remnant}
\author{Y. L.~Zhu}
\email{yzhu14@ncsu.edu}
\affiliation{Department of Physics, North Carolina State University, Raleigh, NC 27695 USA.}
\author{A. Perego}
\email{albino@theorie.ikp.physik.tu-darmstadt.de}
\affiliation{Institut f\"{u}r Kernphysik, Technische Universit\"{a}t Darmstadt, Schlossgartenstr. 2, Darmstadt D-64289,Germany}
\author{G. C.~McLaughlin}
\email{Gail_McLaughlin@ncsu.edu}
\affiliation{Department of Physics, North Carolina State University, Raleigh, NC 27695 USA.}

\date{\today}
\begin{abstract}
The Matter-Neutrino Resonance (MNR) phenomenon has the potential to significantly alter the flavor content of neutrinos emitted from compact object mergers. 
We present the first calculations of MNR transitions  using neutrino self interaction potentials and matter potentials  generated self-consistently from a dynamical model of a three-dimensional neutron star merger.  In the context of the single angle approximation, we find that Symmetric and Standard MNR transitions occur in both normal and inverted hierarchy scenarios.  We examine the spatial regions above the merger remnant where propagating neutrinos will encounter the matter neutrino resonance and find that a significant fraction of the neutrinos are likely to undergo MNR transitions. 

\end{abstract}
\medskip
\maketitle


\section{Introduction}

Remnants arising from binary neutron star mergers have been found to produce immense numbers of neutrinos and antineutrinos. The neutrinos and antineutrinos play important roles in the physics of several phenomena associated with these remnants including jet production in gamma ray bursts \cite{Eichler:1989ve,Ruffert:1998qg,Rosswog:2003rv,Just:2015dba} 
as well as nucleosynthesis from collisionally heated material \cite{Wanajo:2014wha,Sekiguchi:2015dma,Goriely:2015fqa,Roberts:2016igt} as well as winds \cite{Surman:2003qt,Surman:2004sy,Surman:2005kf,Dessart:2008zd,Fernandez:2013tya,Just:2014fka,Martin:2015hxa}.  Neutrino physics is a necessary component in simulations that predict
gravitational radiation from mergers \cite{Sekiguchi:2011zd,Foucart:2014nda,Palenzuela:2015dqa,Bernuzzi:2015opx,Foucart:2015gaa} and additionally, if a merger occurs within range of current or future neutrino detectors, the neutrino signal will provide a wealth of information about these objects \cite{Caballero:2009ww,Caballero:2015cpa}.  An understanding of the  flavor content of the neutrinos is essential for providing an accurate picture of merger phenomena. 

Neutrino oscillations  will occur in neutron star mergers and have the effect of changing the flavor content of the neutrinos. Merger remnants are a dense environment, so flavor transformation is strongly influenced by interactions between neutrinos and the surrounding particles. The potential that a neutrino experiences due to coherent forward scattering on other neutrinos and antineutrinos, sometimes called the neutrino self interaction potential, is a significant driver of the neutrino evolution.  The influence of this potential on supernova neutrino flavor transformation has been extensively studied, e.g. \cite{Duan:2006an,Hannestad:2006nj,EstebanPretel:2008ni, Duan:2010bg,Pehlivan:2011hp, Volpe:2013jgr,Balantekin:2006tg,Vlasenko:2014bva,Duan:2010af,Cherry:2012zw,Gava:2009pj} and references therein.  One of the major consequences of the neutrino self interaction potential in supernovae is the  prediction of a pendulum-like oscillation referred to as a nutation/bipolar oscillation. This type of oscillation is also expected to occur for at least some merger neutrinos \cite{Dasgupta:2008cu,Malkus:2015mda}.

In merger remnant environments, it has been suggested that neutrinos can undergo not only the same type of flavor transformations as in supernova scenarios, but also a novel type of transition called {\it Matter-Neutrino Resonance} (MNR) transitions, first observed in \cite{Malkus:2012ts}.
This phenomenon is distinct from the nutation/bipolar oscillation and typically occurs closer to the neutrino emitting surface.  In both hierarchies MNR transitions can dramatically change neutrino flavor content.  During the transformation the 
neutrinos stay \lq\lq on-resonance" meaning that the neutrinos evolve in such a way so as to ensure that all entries of the Hamiltonian remain relatively small \cite{Malkus:2014iqa}. In addition,  the splitting of the instantaneous eigenstates remains small  \cite{Vaananen:2015hfa}, and the
neutrinos stay approximately on their instantaneous mass eigenstates  \cite{Vaananen:2015hfa,Wu:2015fga}.  The MNR transformation phenomenon requires a matter potential (from neutrino coherent forward scattering on baryons and charged leptons)  and a neutrino self interaction potential of opposite sign and roughly equal magnitude. In merger remnants this condition can be fulfilled either  (1) because  leptonization requires that antineutrinos outnumber neutrinos initially or (2) because the 
geometry of the system causes the relative contributions of neutrino and antineutrinos to the self interaction potential to shift as the neutrinos travel along their trajectories.  This geometric effect comes from  the extension of the neutrino emitting surface  significantly beyond the antineutrino emitting surface, for a discussion see \cite{Malkus:2015mda} 
The matter neutrino resonance is not naively expected to occur in supernovae because both potentials have the same sign.  However, if Non Standard Interactions (NSIs) exist with significant strength 
they can trigger MNR transitions in supernovae \cite{Stapleford:2016jgz}.  Further, if neutrino-antineutrino oscillations occur near the surface of the protoneutron star, a situation similar to Matter Neutrino Resonance transformation occurs  \cite{Vlasenko:2014bva}.

Two primary types of Matter Neutrino Resonance transformation have been suggested to occur in merger remnants.  One of these, the 
{\it Standard} MNR transition, is characterized by a full conversion of electron neutrinos to other types while electron antineutrinos partially transform but then return to their original configuration  ~\cite{Malkus:2014iqa}. 
Standard MNR transitions occur in regions where the neutrino self interaction potential begins as the  largest potential in the system, but declines in magnitude eventually reaching the same magnitude as the matter potential.   
The other type of MNR  transformation fully converts both electron neutrinos and antineutrinos symmetrically in short range and is called a  {\it Symmetric} MNR transition~\cite{Malkus:2015mda}. 
Symmetric MNR transitions differ from Standard MNR transitions in that they occur when geometric effects cause the system to pass from a region where antineutrinos dominate the neutrino self interaction potential to a region where neutrinos dominate this potential \cite{Malkus:2015mda,Vaananen:2015hfa}.  Both types of transformation may well influence nucleosynthesis \cite{Malkus:2012ts,Malkus:2015mda} as MNR transitions typically occur closer to the emitting surface of neutrinos than other large scale oscillation phenomena. 

Previous studies of Matter Neutrino Resonance transitions used phenomenological, flat neutrino emission surfaces motivated by the qualitative properties found in complex dynamical merger simulations. Here, we consider a snapshot of a dynamical calculation \cite{Perego:2014fma} and compute flavor transformation using 
the neutrino self interaction potentials and matter potentials obtained from the same dynamical simulation.  The purpose of this work is to explore the character of matter neutrino resonance transformations in the presence of a  more complex neutrino emission geometry as well as self-consistent density profiles.

This manuscript is structured as follows. In Sec. \ref{secModel} we describe the dynamical calculation and discuss the behavior of the physical quantities relevant for neutrino flavor transformation. In Sec. \ref{secCalculation}, we explain the Hamiltonian and the evolution equations used in our calculations.
In Sec. \ref{secResults}, we present the results of our neutrino flavor transformation calculations and locate the spatial regions where propagating neutrinos will encounter MNRs.
In Sec. \ref{secDiscuss}, we show that MNR transitions are approximately independent of hierarchy although the efficacy of the resonance is determined in part by neutrino mixing angle and mass squared differences. We also consider three flavor effects by comparing two and three flavor scenarios. We conclude in Sec. \ref{secConclude}.


\section{The Merger Remnant}
\label{secModel} 
 
We consider the numerical results of a detailed three dimensional, Newtonian simulation of the aftermath of a binary neutron star merger under the influence of neutrino cooling and heating \cite{Perego:2014fma}.  We choose a merger remnant configuration at $t_0\approx$~40 ms after the beginning of the simulation, corresponding to $\sim$~60 ms after the neutron star merger. At this time the remnant is characterized by an almost axisymmetric, stationary rotating massive neutron star (MNS)
with a baryonic mass of $\sim 2.6 \, {\rm M}_{\odot}$, surrounded by a thick accretion disk with a mass of $\sim 0.15 \, {\rm M}_{\odot}$. The hot and dense matter in the central object and in the disk emits copiously neutrinos of all flavors, with a total luminosity in excess of $10^{53} \, {\rm erg \, s^{-1} }$. 
The re-absorption of a fraction of the emitted radiation inside the disk causes the formation of a neutrino-driven wind that emerges and expands parallel to the rotation axis, on a timescale of a few tens of milliseconds.

  \begin{figure}[ht!]
      \centering
       \centering
    	\includegraphics[angle=0, width=0.7\textwidth]{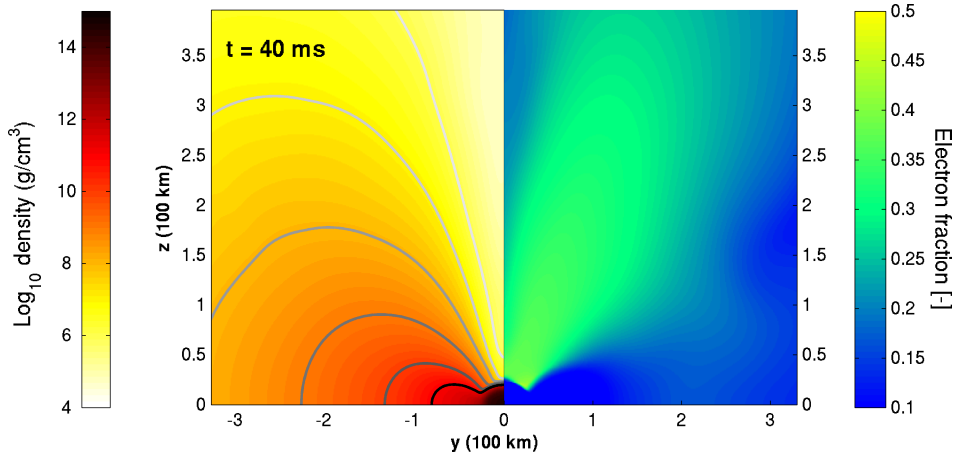}
	\caption{Profiles of the matter density (left) and electron fraction (right) inside the disk and the wind, used as input data for our calculations. Gray lines in the density panel represent 6 logarithmically spaced isodensity contours between $10^6 \, {\rm g \, cm^{-3}}$ (light gray) and $10^{11} \, {\rm g \, cm^{-3}}$ (dark gray).}. 
	\label{fig:profiles}
\end{figure}

The matter inside the wind is characterized by a smooth density profile, slowly decaying from $10^{10} \, {\rm g \, cm^{-3}}$ to the atmosphere ($\lesssim 10^{4} \, {\rm g \, cm^{-3}}$) on a lengthscale of a few hundreds of km.  Due to the strong absorption of $\nu_e$s on neutron rich matter, the electron fraction, $Y_e$, is in the range of 0.25 - 0.4.  The distribution of density and electron fraction above the remnant can be seen in Fig.~\ref{fig:profiles}.  Since the wind develops mainly from the disk, the funnel above the central neutron star has a lower density, while the material extending from the disk in the equatorial direction has high density.  Because of the high degree of axial symmetry in the remnant and in the wind, we compute cylindrical averages for the density.

\begin{figure}[ht!]
	\centering
	\includegraphics[angle=270, width=0.6\textwidth]{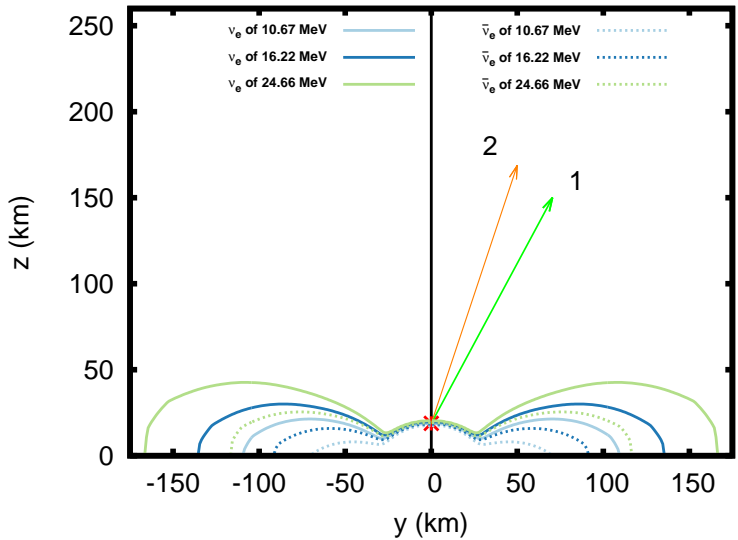}
	\caption{Solid(dashed) lines are neutrino(anti-neutrino) surfaces for energies of 10.67 MeV, 16.22 MeV, 24.66 MeV; green arrow: trajectory 1, test neutrino originating from Cartesian coordinates of (0,0,18) km with a direction of $ (0, \sqrt{2}, 3 )$ ;  orange arrow: trajectory 2, test neutrino from (0,0,18) km with a direction of (0,1,3). 
    }. 
	\label{fig:tra1_2}
\end{figure}

 The neutrino emission rates and opacities are computed using an energy dependent (spectral)  leakage scheme, whose details can be seen in Ref.~\cite{Perego:2014fma,Perego:2015agy}.
 The neutrino reactions included in the calculations are the emission, absorption and scattering off free nucleons. Neutrino pair emission from electron-positron annihilation and nucleon-nucleon bremsstrahlung are also included, as well as absorption from their inverse reactions.  
 Neutrino type is distinguished as $\nu_e$, $\bar{\nu}_e$ and $\nu_x$, the latter being a collective species  for 
 $\mu$ and $\tau$ neutrinos and antineutrinos.
 Energy dependent neutrino optical depths are computed by minimizing the line  integral of the inverse mean free path along several possible propagation directions.  We distinguish between scattering and energy optical depths. In the former case, the neutrino surfaces correspond to the last scattering surfaces. In the latter case, the neutrino surfaces denote the transition from a thermally to a non-thermally coupled regime.
 Neutrinos diffusing from the optically thick neutron star and inner disk are considered 
 emitted isotropically from neutrino (last scattering) surfaces, while neutrinos produced in optically thin conditions
 are considered emitted isotropically from their production site.  The values of the emitted neutrino mean energies 
 remain similar during the simulation 
 and at $t_0 \approx 40$ ms they are $\langle E_{\nu_e} \rangle  \approx10.6 $ MeV, $\langle E_{\bar{\nu}_e} \rangle \approx15.3$ MeV and $\langle E_{\nu_x} \rangle \approx17.3 $ MeV. 
 Neutrino luminosities decrease steadily, but slowly, during the first hundreds ms. At $t=t_0$, the integrated $\bar{\nu}_e$ luminosity measured immediately above the most relevant neutrino surfaces is $\sim 4 \times 10^{52}~{\rm erg\,s^{-1}}$, while the $\nu_e$ luminosity is $\sim 3 \times 10^{52}~{\rm erg\,s^{-1}}$. For each heavy flavor neutrino, we have $L_{\nu_x} = L_{\nu_x,{\rm total}}/4   \sim 9 \times 10^{51}~{\rm erg\,s^{-1}}$.  Neutrino quantities are binned logarithmically in energy and a few examples of electron neutrino and antineutrino scattering surfaces of different energy levels are shown in Fig.~\ref{fig:tra1_2}, along with a couple example neutrino trajectories that will be discussed in the next sections. 
 
The local radiation intensity for neutrinos  everywhere above the emission surface for each species of neutrino  is given by $I_{\nu_{\alpha}}$.  The quantity  $I_{\nu_{\alpha}}$ is defined such that 
\begin{equation}
n_{\nu_{\alpha}}(\mathbf{x},t) = \int_0^{\infty} \int_\Omega \frac{I_{\nu_{\alpha}} ( E,\mathbf{n}',\mathbf{x},t )}{E} \, d \Omega \, d E
\end{equation}
is the local density of neutrino $\nu_{\alpha}$ at $\mathbf{x}$ and time $t$, and $dN_{\nu_{\alpha}}(\mathbf{x},t) = I_{\nu_{\alpha}} \left( E,\mathbf{n}',\mathbf{x},t \right) /E \; d\Omega \, dE \, dV$ 
 is the number of neutrinos $\nu_{\alpha}$ with energy in an interval $E \pm dE/2$, contained in a volume element ${\rm d}V$ around $\mathbf{x}$, that moves
 inside a cone of amplitude ${\rm d}\Omega$, of vertex $\mathbf{x}$ and axis along $\mathbf{n}'$. The quantity $I_{\nu_{\alpha}}$ is computed from the spectral emission rates using a ray-tracing
algorithm. The ray tracing calculation produces a pattern of fluxes above the remnant that is both non-spherical and  different for the $\nu_e$ and the $\bar{\nu}_e$.  This pattern is influenced by the a couple of key factors.

 First, a fraction of the neutrinos that are emitted at the neutrino surfaces get absorbed on their way out of the object.   
 We take into account the  absorption of neutrinos along a propagation ray by multiplying the radiation intensity by $\exp(-\Delta \tau_{\rm en})$, where $\Delta \tau_{\rm en}$ is the difference  in the energy optical depths due to absorption processes between the emission point and any point along the ray.  This absorption is a spatially dependent effect.  It effects neutrinos and antineutrinos from the disk more than those from the massive neutron star because the region above the massive neutron star is less dense than that above the disk.  Moreover,  the absorption affects neutrinos and antineutrinos at a different level.  The $\nu_e$s  are more easily absorbed than $\bar{\nu}_e$s because there are more neutrons than protons in the regions above the emission surfaces. The electron neutrino luminosity can be significantly reduced along escaping paths (up to $\sim 35 \%$).  At infinity, therefore, the integrated electron antineutrino flux is larger than the electron neutrino flux.
 
 Second, the geometry of the MNS and the disk is not conducive to an isotropic flux.
 Different parts of the remnant emit differing amounts of neutrino flux, with a slightly smaller component coming from the MNS than the disk.  Due to the presence of the optically thick disk, the neutrino fluxes are larger along the poles than along the equator.  In addition, the two parts of the remnant emit different relative numbers of $\nu_e$ and $\bar{\nu}_e$.
 On the one hand, the high density ($\rho \gtrsim 10^{12} \, {\rm g \, cm^{-3}} $) and small electron fraction inside the MNS favor the presence of a larger number of trapped $\bar{\nu}_e$s than $\nu_e$s \cite{Foucart:2015gaa}. The former diffuse out on a smaller timescale, due to their lower opacity. Thus, the MNS emits more electron antineutrinos than neutrinos.
On the other hand, the emission from the hot disk, powered by the accretion process, happens mostly at lower densities and higher electron fractions. Under these conditions, a larger number of softer electron neutrinos is emitted from a wider portion of the disk, compared with electron antineutrinos.

\begin{figure}[ht!]
	\centering
    \includegraphics[angle=0, width=0.7\textwidth]{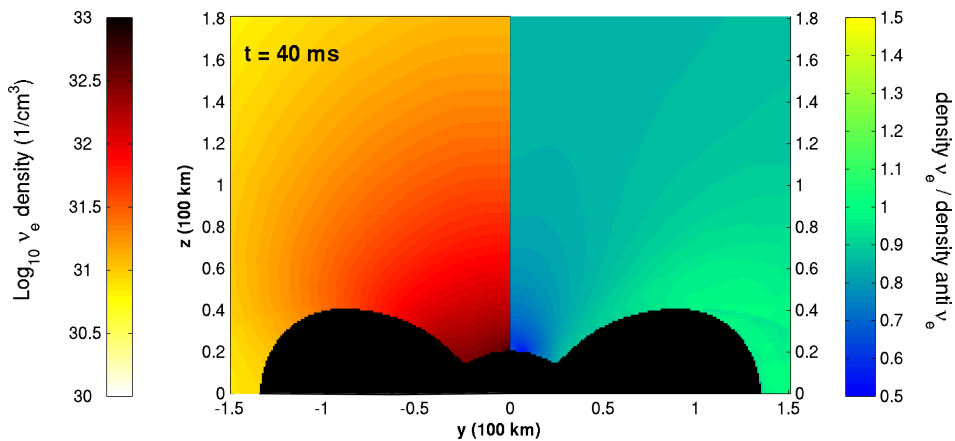}
	\caption{Profiles of the electron neutrino density (left, logarithmic scale) and of the ratio between the electron neutrino and antineutrino densities (right, linear scale) above the outermost neutrino surface; cut is at $\rho=     10^{10} \, {\rm g \, cm^{-3}}$. 
	\label{fig:neutrino densities}}
\end{figure}

The combination of both of these factors  
determines the non-trivial spatial dependence of the neutrino densities above the remnant, which is presented in Fig.~\ref{fig:neutrino densities}. The more intense $\bar{\nu}_e$ fluxes coming from the MNS and the more compact $\bar{\nu}_e$ surfaces lead to the presence of more abundant electron antineutrinos in the funnel above the MNS. At larger distances from the rotational axis, the disk neutrinos play an increasingly important role, altering the ratio of neutrinos and antineutrinos.  The larger absorption provided by the disk on $\nu_e$ smooths the transition between the two different regimes. 

\section{Calculations}
\label{secCalculation}
We are interested in determining the flavor evolution of the neutrinos and anti-neutrinos after they leave the trapped region of the disk/neutron star and propagate outward. 
As the neutrinos travel along their trajectories, the scattering matrices for neutrinos and antineutrinos evolve  
according to the evolution equations
\begin{equation}
\begin{aligned}
{\rm i} \frac{d}{dt} S(E,\mathbf{x},t)&=H(E,\mathbf{x},t) \,S\left(E,\mathbf{x},t\right) ,
\\
{\rm i} \frac{d}{dt} \bar{S}(E,\mathbf{x},t)&=\bar{H}(E,\mathbf{x},t) \,\bar{S}\left(E,\mathbf{x},t\right),
\label{eq:EOS}
\end{aligned}
\end{equation}
where $S(E,\mathbf{x},t)$ and $\bar{S}(E,\mathbf{x},t)$ are the scattering matrices for neutrinos and antineutrinos respectively as discussed in  \cite{Kneller:2009vd}. 
Similarly $H$ and $\bar{H}$ are the Hamiltonians for the neutrinos and antineutrinos. 
For neutrinos, the total Hamiltonian is
\begin{equation}
\label{eq:H}
H(E,\mathbf{x},t)=H_V(E) + H_e(\mathbf{x},t) +  H_{\nu\nu}(\mathbf{x},t).
\end{equation}
We perform most flavor transformation calculations  with two species of neutrinos, electron neutrinos and mu/tau neutrinos i.e. $\nu_e  \leftrightarrow \nu_x$, and the corresponding antineutrinos, $\bar{\nu}_e  \leftrightarrow \bar{\nu}_x$, 
although in Sec. \ref{secDiscuss} we explicitly check this approximation by comparing two and three flavor results. 
 Note that in the dynamical simulation, $\nu_x$ 
 represents the collective species $\nu_\mu$, $\nu_\tau$, $\bar{\nu}_\mu$, $\bar{\nu}_\tau$ but when describing two flavor oscillation calculations in this section $\nu_x$ indicates only one species, most accurately described as a linear combination of $\nu_\mu$ and $\nu_\tau$.

In a two flavor system, the vacuum Hamiltonian $H_V$ is given in the flavor basis by 
\begin{equation}
\label{eq:HV}
H_V(E)=\frac{\Delta_V}{2}\left(
		\begin{array}{cc}
		-\cos2\theta_V & \sin2\theta_V\\
		\sin2\theta_V & \cos2\theta_V
		\end{array}
	\right) \ ,
\end{equation}
where $\Delta_V \equiv \delta m_{ij}^2/(2 E)$, the neutrino mass-squared difference in vacuum $\delta m_{ij}^2=m_i^2-m_j^2$,  vacuum mixing angle $\theta_V$ and neutrino energy, $E$.  For most of the calculations presented here, we use a mixing angle of $\theta_{13}=0.15$ \cite{An:2013zwz} with $\Delta_{13}=2.43\times 10^{-3} \, {\rm eV}^2$ \cite{Agashe:2014kda}
and the normal hierarchy, although we explore the dependence of our results on these parameters in Sec. \ref{secDiscuss}. The matter Hamiltonian $H_e$ can be written as  
\begin{equation}
\label{eq:He}
H_e(\mathbf{x},t) =\left(
		\begin{array}{cc}
			V_e(\mathbf{x},t) & 0 \\
			0 & 0
		\end{array}
	\right),
\end{equation}
where the matter potential $V_e=\sqrt{2} G_F n_e$, $n_e=\rho_m Y_e N_A.$   $G_F$ is the Fermi constant, $n_e$ is the number density of electrons minus number density of positrons, $\rho_m$ the matter density, $Y_e$ is the matter electron fraction and $N_A$ is the Avogadro number.
The neutrinos and anti-neutrinos experience a potential from the ambient neutrinos, or the neutrino-neutrino self-interaction Hamiltonian, which is given by
\begin{equation}
\label{eq:Hnunu}
  H_{\nu\nu}(\mathbf{x},t)=H_\nu(\mathbf{x},t)-H_{\bar{\nu}}(\mathbf{x},t), 
\end{equation}
which has a component from the interactions with neutrinos, and another from the interactions with antineutrinos 
\begin{equation}
\label{eq:hnu}
\begin{aligned}
	H_\nu(\mathbf{x},t)=& \int_0^\infty\left(S(E,\mathbf{x},t) \,  \rho(E,\mathbf{x},t) \, S^\dagger(E,\mathbf{x},t) \right)\,dE \,    , \\ 
		H_{\bar{\nu}}(\mathbf{x},t)=&\int_0^\infty\left(\bar{S}^*(E,\mathbf{x},t)\, \bar{\rho}(E,\mathbf{x},t)\, \bar{S}^T(E,\mathbf{x},t)\right)\,dE .
\end{aligned}
\end{equation}
For each neutrino test trajectory, $\mathbf{x}$ has a prescribed relation to $t$, $\mathbf{x}=\mathbf{x}(t)$.  We use the single angle approximation, meaning that we assume that all ambient neutrinos have the same history as the test neutrino.
Initially at time $t=t_0$ which corresponds to an $\mathbf{x}_0$ for each trajectory, the scattering matrices $S$ and $\bar{S}$ are identity matrices. 
Following Ref. \cite{Malkus:2015mda}, we define the quantities 
$\rho(E,\mathbf{x},t) \, dE$
and 
$\bar{\rho}(E,\mathbf{x},t) \, dE$
as the unoscillated neutrino potential matrices in energy range $E\pm dE/2$, i.e. the values the Hamiltonian would have if the scattering matrices remained as identity matrices. 
At the initial emission point $\mathbf{x_0}$, 
the neutrino has not had the opportunity to flavor transform, so we start with the neutrino-neutrino self-interaction Hamiltonian, 
\begin{equation}
\begin{aligned}
	H_\nu(\mathbf{x_0},t_0)=&\begin{pmatrix}
		H_{\nu,ee}(\mathbf{x_0},t_0) & H_{\nu,ex}(\mathbf{x_0},t_0)  \\
		H_{\nu,xe}(\mathbf{x_0},t_0)  & H_{\nu,xx} (\mathbf{x_0},t_0)
	\end{pmatrix}
=
		\begin{pmatrix}
		V_{\nu_e}(\mathbf{x_0},t_0) & 0 \\
		0 &  V_{\nu_x}(\mathbf{x_0},t_0)
	\end{pmatrix}
	,\\
	H_{\bar{\nu}}(\mathbf{x_0},t_0)=&
		\begin{pmatrix}
		H_{\bar{\nu},ee}(\mathbf{x_0},t_0) & H_{\bar{\nu},ex}(\mathbf{x_0},t_0)  \\
		H_{\bar{\nu},xe}(\mathbf{x_0},t_0)  & H_{\bar{\nu},xx}(\mathbf{x_0},t_0) 
	\end{pmatrix}
=\begin{pmatrix}
			V_{\bar{\nu}_e}(\mathbf{x_0},t_0) & 0 \\
			0 & V_{\bar{\nu}_x}(\mathbf{x_0},t_0)
	\end{pmatrix}.
\end{aligned}\label{eq:HnuV}
\end{equation}
In general, the unoscillated potentials $V^{\rm un}_{\nu_\alpha}$ at location $\mathbf{x}$ due to the weak interaction between a test neutrino moving in
 the direction $\mathbf{n}$ and the ambient neutrinos $\nu_{\alpha}$ are
 \begin{equation}
 V^{\rm un}_{\nu_{\alpha}}(\mathbf{x},t) = \sqrt{2} \, G_{F} \int_0^{\infty}  \:
        \left( \int_{\Omega} \: \frac{I_{\nu_{\alpha}} \left( E, \mathbf{n}', \mathbf{x}, t \right)}{E}  \left( 1 - \cos{\Theta} \right) \: d \Omega  \right) \: dE \,.\label{eq:neuPot}
 \end{equation}
 where  $\Omega$ is the solid angle associated with the local neutrino directions $\mathbf{n}'$, 
 $\Theta$ is the angle between $\mathbf{n}'$ and $\mathbf{n}$. 
 So initially, $V_{\nu_\alpha}(\mathbf{x_0},t_0) =  V^{\rm un}_{\nu_{\alpha}}(\mathbf{x_0},t_0)$, although as the  system evolves, $H_\nu(\mathbf{x},t)$ and $H_{\bar{\nu}}(\mathbf{x},t)$ pick up off diagonal components 
 and the $V_{\nu_\alpha(\bar{\nu}_\alpha)}$ 
are in general modified by the scattering matrices as in Eq.~\ref{eq:hnu}.

We can adjust the form of the Hamiltonian by subtracting its trace without impacting the flavor transformation results. In this case  the diagonal terms of total Hamiltonian become
\begin{equation}
\begin{aligned}
\label{eq:Hdiaw1gonal}
H(E,\mathbf{x},t)_{ee} = \frac{1}{2}(-\Delta_V \cos2\theta_V + V_e(\mathbf{x},t) + V_{\nu}(\mathbf{x},t)) ,
\\
H(E,\mathbf{x},t)_{xx} = \frac{1}{2}(\Delta_V \cos2\theta_V -( V_e(\mathbf{x},t) + V_{\nu}(\mathbf{x},t))) ,
\end{aligned}
\end{equation}
where 
\begin{equation}
\label{eq:Vnu}
	\begin{aligned}
		V_{\nu}(\mathbf{x},t) &\equiv V_{\nu_e}(\mathbf{x},t)-V_{\nu_x}(\mathbf{x},t)-\left(V_{\bar{\nu}_e}(\mathbf{x},t)- V_{\bar{\nu}_x}(\mathbf{x},t)\right)\ .
		 \\
	\end{aligned}
\end{equation}
The antineutrino Hamiltonian $\bar{H}(E,\mathbf{x},t)=-H_V(E) + H_e(\mathbf{x},t) +  H^*_{\nu\nu}(\mathbf{x},t)$
can be obtained in the same manner. 

We will use Eq. \ref{eq:Vnu} and its unoscillated version, i.e. the potential if no oscillation were to occur,
\begin{equation}
\label{eq:Vnun}
	\begin{aligned}
		V^{\rm un}_{\nu}(\mathbf{x},t) &\equiv V^{\rm un}_{\nu_e}(\mathbf{x},t)-V^{\rm un}_{\nu_x}(\mathbf{x},t)-\left(V^{\rm un}_{\bar{\nu}_e}(\mathbf{x},t)- V^{\rm un}_{\bar{\nu}_x}(\mathbf{x},t)\right)\ .
		 \\
	\end{aligned}
\end{equation}
in order to discuss the behavior of the matter neutrino resonance transitions in the next sections.  We also use energy dependent survival probabilities defined in terms of the scattering matrices as $P_{\nu_e \rightarrow \nu_e}(E, \mathbf{x},t) = |S_{e e} (E, \mathbf{x},t) |^2$ and $P_{\bar{\nu}_e \rightarrow \bar{\nu}_e} (E, \mathbf{x},t) = |\bar{S}_{e e} (E, \mathbf{x},t) |^2$.  If the survival probability is one, no flavor transformation has occurred.  However, if the survival probability is zero, complete transformation has occurred, so the neutrinos in that species have completely transformed. 


\section{Results}
\label{secResults}

We now discuss the results of the multi-energy numerical evolution of the neutrino and antineutrino scattering matrices for a number of trajectories in the context of the  binary neutron star merger simulation described in Sec. \ref{secModel}.
We assume that the matter and the neutrino profiles taken from the simulation at $t=t_0$ are stationary over the propagation timescale of the test neutrinos. Thus, the temporal dependence is only contained inside the trajectory equation $\mathbf{x}=\mathbf{x}(t)$.
When presenting results for single neutrino trajectories, we follow Ref. \cite{Malkus:2015mda} and plot weighted survival probabilities of electron neutrinos and anti-neutrinos,
\begin{align}
P_{\rm \nu_e,num} (\mathbf{x}) & = \left< P_{\nu_e \rightarrow \nu_e} (E,\mathbf{x}) \right> = \frac{1}{n_{\nu_e}(\mathbf{x},t_0)} \int^{\infty}_0 \left( \int_{\Omega} \frac{I_{\nu_e}(E,\mathbf{n}',\mathbf{x},t_0)}{E} \: d\Omega \right) 
P_{\nu_e \rightarrow \nu_e}(E,\mathbf{x}) \: dE , \\
P_{\rm \bar{\nu}_e,num} (\mathbf{x}) & = \left< P_{\bar{\nu}_e \rightarrow \bar{\nu}_e} (E,\mathbf{x}) \right> = \frac{1}{n_{\bar{\nu}_e}(\mathbf{x},t_0)} \int^{\infty}_0 \left( \int_{\Omega} \frac{I_{\bar{\nu}_e}(E,\mathbf{n}',\mathbf{x},t_0)}{E} \: d\Omega \right) 
P_{\bar{\nu}_e \rightarrow \bar{\nu}_e}(E,\mathbf{x}) \:  dE
\label{eq:wp}
\end{align}
Since we begin with neutrinos in flavor states and focus on the matter neutrino resonance region, generally flux averaged survival probability results have converged with only sixteen energy bins, although we check the convergence with calculations of one thousand energy bins.

We first consider neutrinos that are emitted from the massive neutron star, since a significant fraction of neutrinos are emitted from this object,  As an example, we examine neutrinos which are emitted at the $10.67$ MeV electron neutrino surface (solid cyan line in Fig.~\ref{fig:tra1_2}), considering two trajectories which correspond to arrows one and two in Fig.~\ref{fig:tra1_2}.  Trajectory one has
an angle of approximately 25 degrees to the vertical and trajectory two has an angle of approximately 18 degrees to the vertical. 

\begin{figure}[ht!]
    \centering
    \begin{subfigure}
            \centering
        \includegraphics[angle=270, width=3.4in]{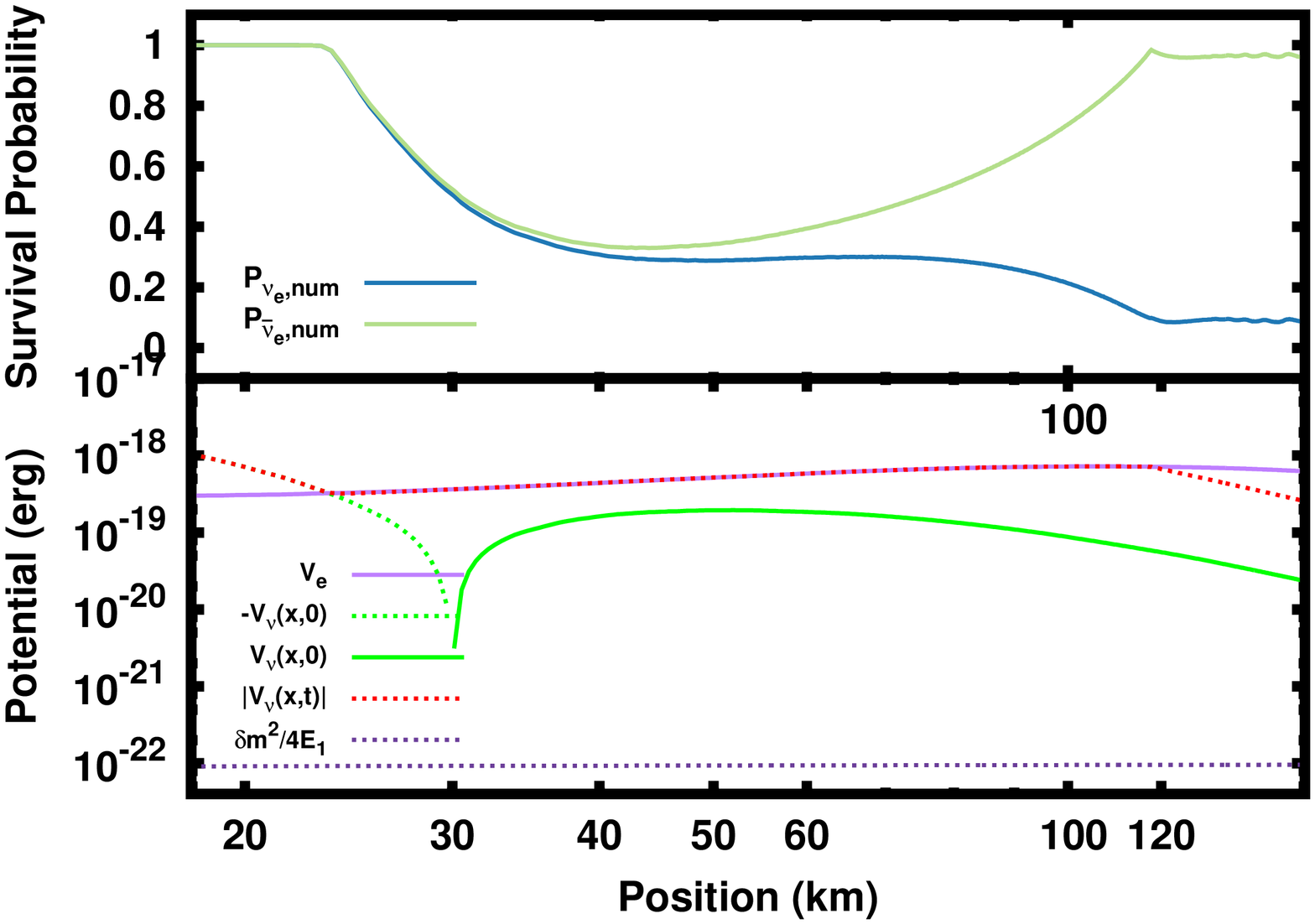}
            \end{subfigure}%
    ~ 
    \begin{subfigure}
        \centering
        \includegraphics[angle=270, width=3.4in]{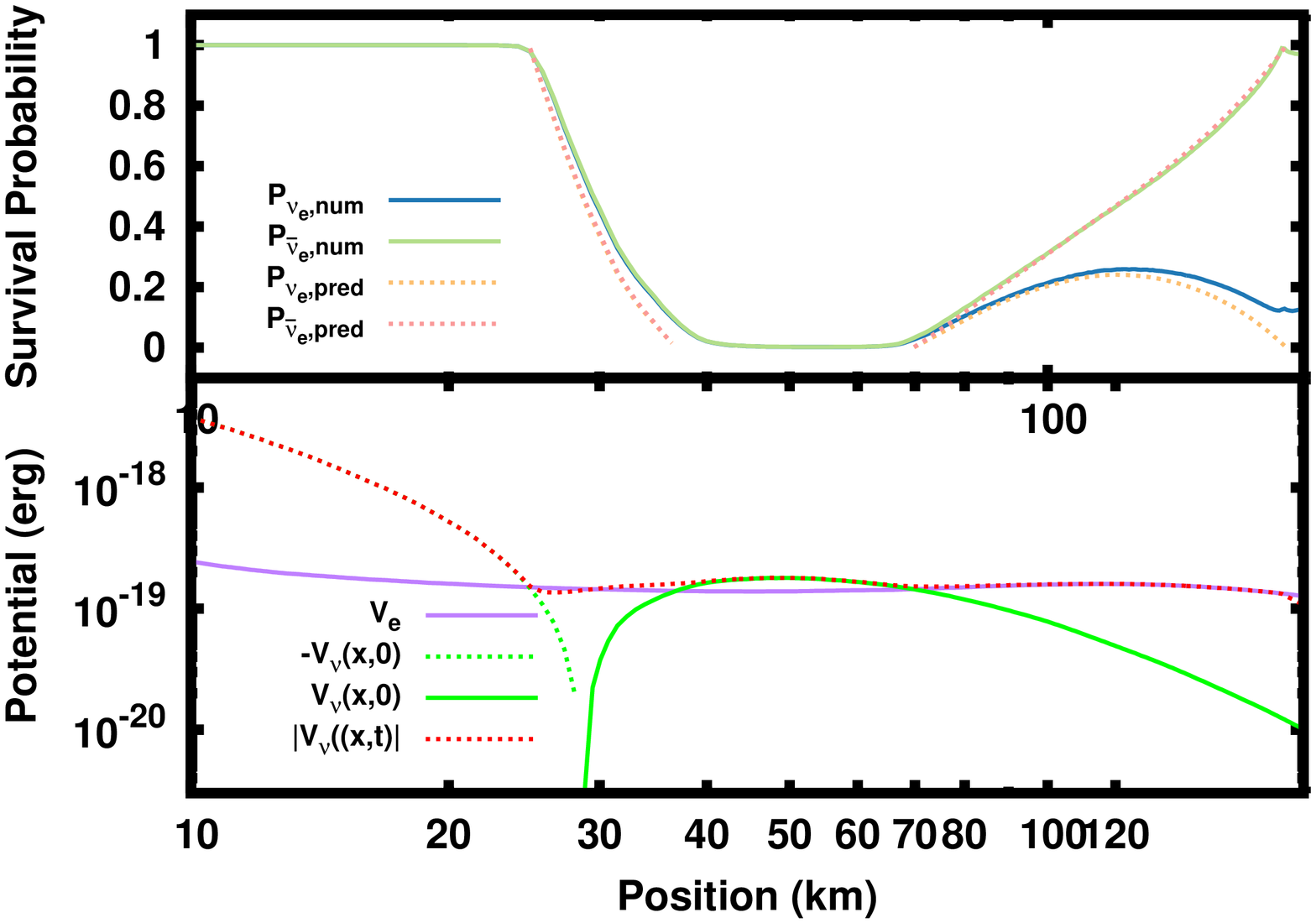}
    \end{subfigure}
    \caption{Top figures: flux weighted numerical survival probability (solid)  as a function of distance along the trajectory for trajectories one (left) and two (right).  Also shown in the top right figure is the predicted survival probability for MNR transitions (dashed). The bottom figures show corresponding potentials: Matter potential (purple) $V_e(\mathbf{x})$, magnitude of the oscillated self-interaction potential (dashed red) $V_\nu(\mathbf{x})$, positive (dashed green) and negative (solid green) unoscillated self-interaction potential, $V^{\rm un}_\nu(\mathbf{x})$, vacuum contribution (dashed purple). Matter neutrino resonance transitions can be seen in the evolution of both trajectories.
\label{fig:16wp12_13}}
\end{figure}
Initially, the neutrinos along these trajectories encounter large negative self interaction potentials which are shown as the dashed red lines in the lower panels of 
Fig.~\ref{fig:16wp12_13}.  This suppresses the flux weighted survival probabilities for both neutrinos and antineutrinos which are shown hovering near one in the top panels of Fig.~\ref{fig:16wp12_13}.
Previous works \cite{Malkus:2012ts,Malkus:2014iqa, Malkus:2015mda, Vaananen:2015hfa,Wu:2015fga} have 
shown that a MNR transition is triggered at roughly the point where the total on diagonal Hamiltonian becomes approximately zero, i.e.
\begin{equation}
	V_e(\mathbf{x}) + V_{\nu}(\mathbf{x}) \approx 0\ .
	\label{mnrcondition}
\end{equation}   
We find the place along the trajectory where Eq. \ref{mnrcondition} is satisfied assuming that no transformation has taken place yet (survival probabilities of one are used to determine $V_{\nu}(\mathbf{x})$) and take this point as the prediction of the starting point of the transition. This predicts the beginning of the transition to be at 24 km and can be seen in the plot as the point
where the dashed red line first intersects with the solid purple line.  During  matter neutrino resonance transitions the neutrinos and antineutrinos transform in such a way as to stay \lq\lq on resonance", i.e. $V_{\nu}(\mathbf{x})$ changes in such a way as to keep Eq. \ref{mnrcondition} fulfilled. In the bottom panels, $V_{\nu}(\mathbf{x})$ (dashed red line) can be seen tracking $V_e(\mathbf{x})$ (solid purple line) during MNR transitions. Matter neutrino resonance transitions end when these lines diverge. 

The transition in trajectory one (left panel) is characterized by electron neutrino and antineutrino survival probabilities that are  
$ P_{\nu_e,{\rm num}}  \approx 0$ and $P_{\bar{\nu}_e,{\rm num}}  \approx 1$ at the end of the transition, making the behavior of this transition most consistent with
a Standard MNR. To locate the end of the transition, we follow ~\cite{Malkus:2014iqa}, and again use Eq. \ref{mnrcondition}, finding the place where it is satisfied for these survival probabilities.  This procedure predicts that the transition will end at  118 km, again in agreement with the numerical results. Note that for trajectory one (left panel) the matter potential is increasing because the neutrino starts in the low density funnel above the center of the massive neutron star but drifts toward the edge.

Trajectory two (right panel) has a somewhat different behavior.  While the unoscillated neutrino interaction potential is similar to that of trajectory one, the matter potential is  nearly constant
as the neutrino remains close to the edge of the low density funnel.  It can be seen in the bottom right panel that matter potential crosses the (magnitude of the) unoscillated neutrino potential three times.  The first two crossings bracket a Symmetric MNR.  Consistent with a Symmetric MNR transitions, both neutrino and antineutrino survival probabilities go from one to zero, between about 24 km and around 40 km, and the unoscillated neutrino self interaction potential goes from positive to negative (dashed green line becomes a solid green line).  
This change in sign is a combination of the properties of the neutrino densities (see Fig.~\ref{fig:neutrino densities}) and a geometric effect. The ratio of the $\nu_e$ to $\bar{\nu}_e$ number density increases to nearly one along the trajectory but the ratio
of the potentials surpasses one because
the neutrino emitting surface is larger than the antineutrino emitting surface and the potentials are weighted by the angle of incidence of the test neutrino to the ambient neutrinos \cite{Malkus:2015mda}.

Comparing with trajectory one, we see there that this change in sign of the unoscillated neutrino potential also occurs, but there is no second crossing of the neutrino self interaction potential with the matter potential, so the complete flavor swap does not occur in the antineutrino channel.  However, one can see the beginnings of Symmetric MNR  as the survival probabilities for neutrinos and antineutrinos initially track each other.  Trajectory one might also be considered as a hybrid MNR, i.e. a part Symmetric, part Standard MNR, instead of a purely Standard MNR. 

Returning to trajectory two, we see that after the Symmetric MNR, the two potentials no longer track each other between around 40 km and 70 km (bottom right panel of Fig.~\ref{fig:16wp12_13}).  This is because for this situation, MNR 
flavor transformation will not allow the neutrino self interaction potential to match the matter potential at the same time it fulfills the other MNR conditions, for example requiring that the off diagonal components of the Hamiltonian remain small \cite{Malkus:2014iqa}.  In this region, no observable flavor transformation is occurring, and both survival probabilities remain at zero. Later at around 70 km, the third crossing occurs.  At this time, the neutrinos begin transforming again, exhibiting the features of a Standard MNR transition.  

We next consider the analytic prediction of survival probabilities during a MNR transition for a single energy system which are given in ~\cite{Malkus:2014iqa,Vaananen:2015hfa}. Neglecting the vacuum scale,
\begin{equation}
\label{eq:anal}
\begin{aligned}
P_{\nu_e,{\rm pred}} &=  \frac{1}{2} \left(1 + \frac{\alpha^2 - R^2 - 1}{2 R} \right)  \ , \\
P_{\bar{\nu}_e,{\rm pred}} &= \frac{1}{2} \left(1 + \frac{\alpha^2 + R^2 - 1}{2 \alpha R} \right) \ ,
\end{aligned}
\end{equation}
where $R \equiv V_e(\mathbf{x})/V^{\rm un}_{\bar{\nu}_e}(\mathbf{x})$ is the ratio of the neutrino-electron and neutrino-neutrino interaction scales as in Eq.~\ref{eq:He} and~\ref{eq:Vnun}, and the asymmetry $\alpha$ is the ratio of $V^{\rm un}_{\bar{\nu}_e}(\mathbf{x})$ to $V^{\rm un}_{\nu_e}(\mathbf{x})$.  We plot this prediction of the survival probabilities for electron neutrinos and antineutrinos during MNR transitions in the top right panel of Fig.~\ref{fig:16wp12_13} as the yellow and orange dashed lines respectively.  The predicted survival probabilities track the numerical results closely during the majority of each transition.

\begin{figure}[ht!]
\centering
    \begin{subfigure}
            \centering
        \includegraphics[angle=270, width=3.4in]{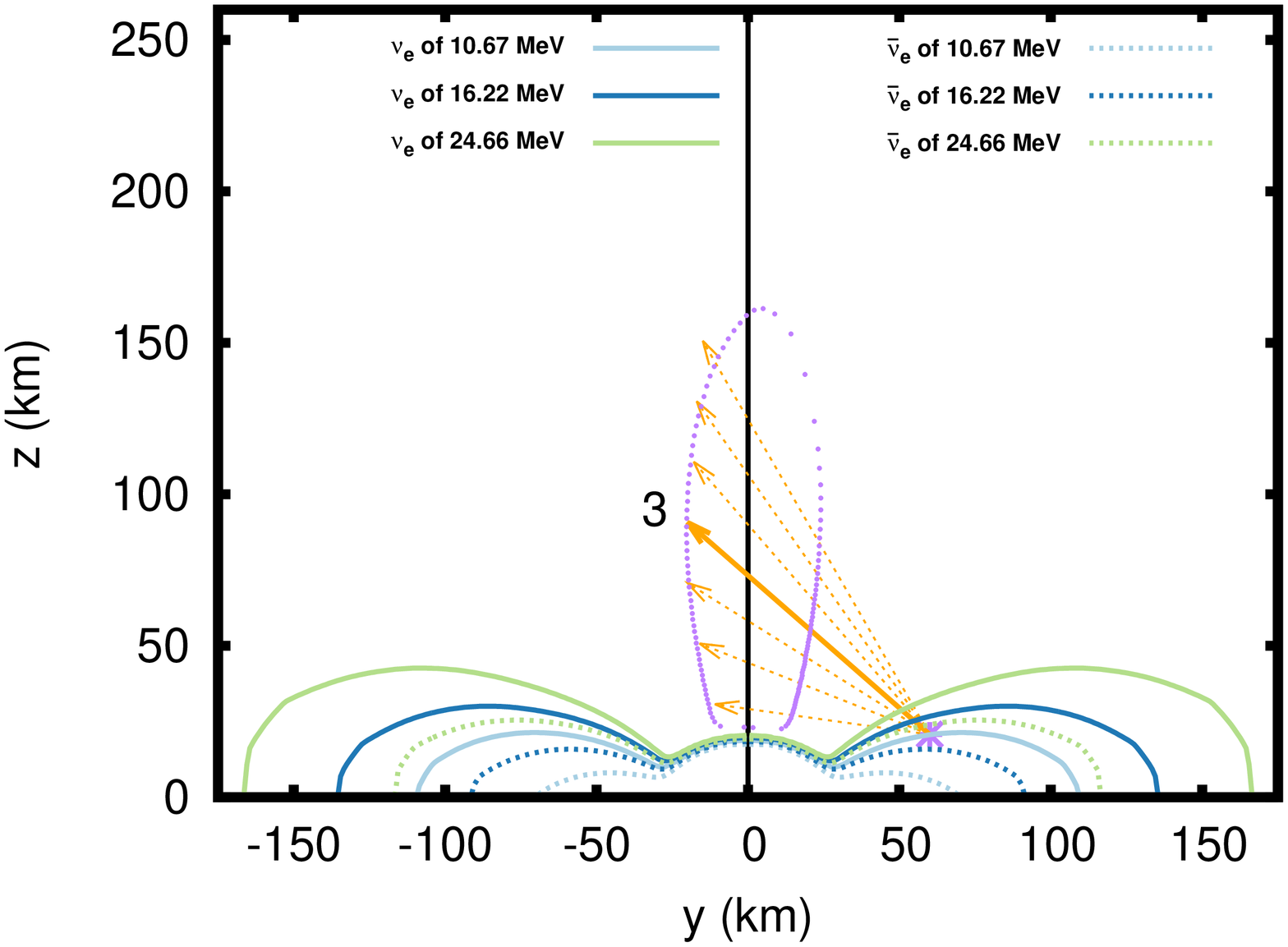}
            \end{subfigure}
    ~ 
    \begin{subfigure}
        \centering
        \includegraphics[angle=270, width=3.4in]{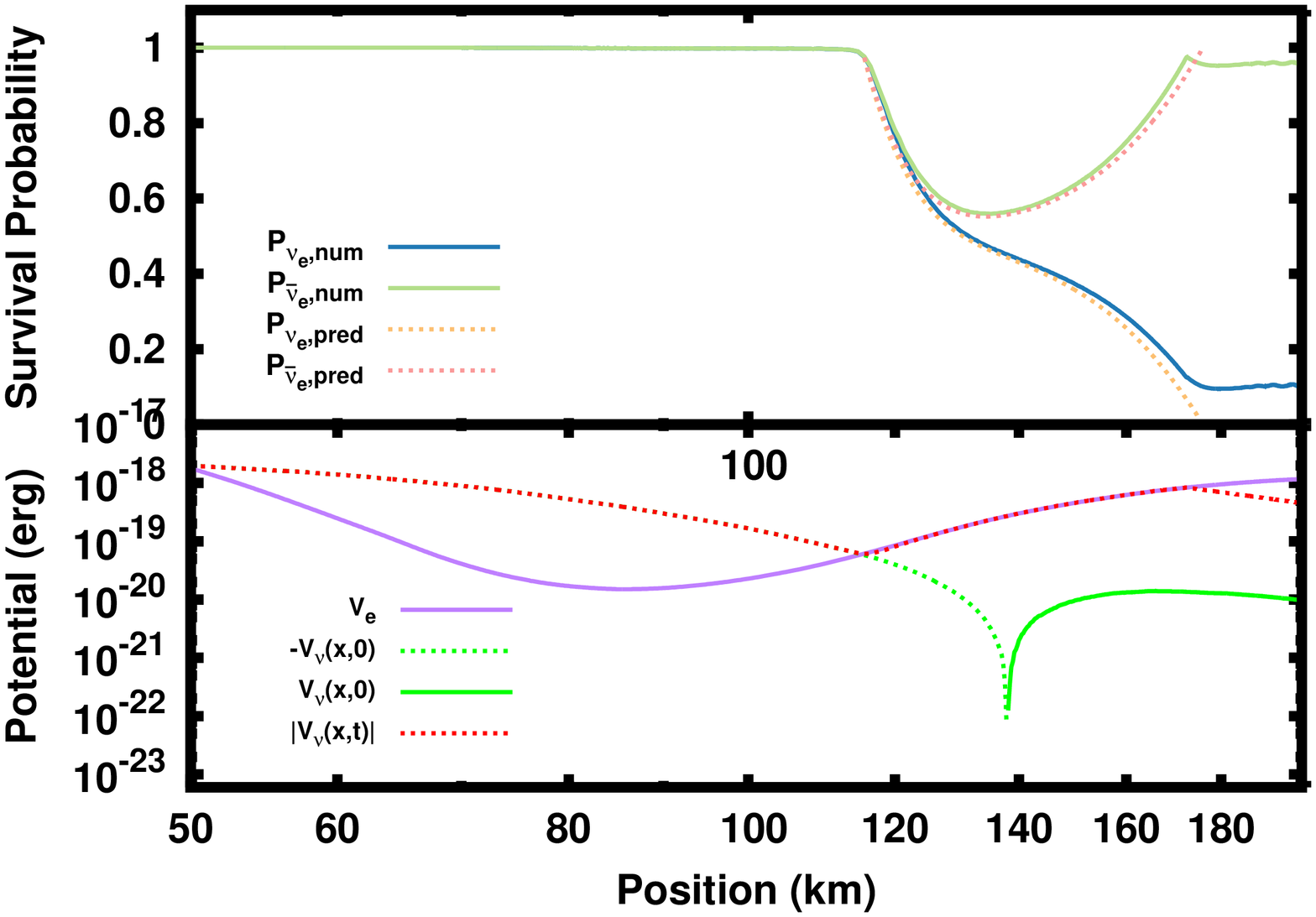}
    \end{subfigure}

        \caption{Left panel:  The orange arrow represents trajectory three, which originates at (0, 60.05, 20.69) km (purple star) and has a direction of (0,-1,1). 
The dotted purple circle above the massive neutron star indicates the MNR surface, defined as the places where 
$V_e+ V_{\nu}^{\rm un} \approx 0 $, for the set of neutrino trajectories originating at the purple star. Neutrino surfaces are shown as in Fig.~\ref{fig:tra1_2}. Right panel: as in Fig.~\ref{fig:16wp12_13} but for trajectory three.} \label{fig:tra3}
\end{figure}

Having considered two neutrino trajectories that begin at the neutron star, we now consider trajectories that originate from the accretion disk. We begin with trajectory three, which is shown in the left panel of  Fig.~\ref{fig:tra3}.  This trajectory travels at an angle of 45 degrees to the vertical toward   
the rotational axis of the system.
In the right panel we show the results of the flavor transformation calculation for this trajectory.  Looking at the lower panel we see that the matter potential (purple solid line) begins high, and then dips.  This dip occurs as the neutrino passes through the underdense funnel over the massive neutron star.  As expected, the Matter Neutrino Resonance transition begins when the matter potential intersects the neutrino self interaction potential.  This is another case of a hybrid (part Symmetric/ part Standard) MNR transition.  The survival probabilities  for electron neutrinos and antineutrinos (upper right panel) are at first quite similar.  However, after the sign change in the potential, the oscillation continues as a Standard MNR.   We see again that the analytic prediction Eq. \ref{eq:anal} matches well the numerical results.  Finally, we note that for this trajectory there is a very early crossing, at about 50 km, which does not result in a MNR transition.  Since the matter potential begins above the neutrino self interaction potential, matter neutrino resonance transitions are not predicted to occur.  At this crossing, the neutrinos cannot fulfill all of the required  conditions for the matter neutrino resonance transitions ~\cite{Malkus:2014iqa,Vaananen:2015hfa}.

We examine a series of additional neutrino trajectories emitted from the same point on the disk.  These are shown as the dotted lines with arrows in the left panel of Fig.~\ref{fig:tra3}.  Purple dots are the locations where for each trajectory, $V_e(\mathbf{x},t) + V^{\rm un}_{\nu}(\mathbf{x},t) \approx 0$.  In this way we locate the MNR surface for that emission point, i.e. the spatial regions above the compact object merger where a MNR transition may occur.  The left and right edges of this surface bracket the low density funnel.  As just discussed in the case of trajectory one, transformation only happens as the neutrinos exit this funnel region where the matter potential is rising above the neutrino self interaction potential.  

\begin{figure}[ht!]
	\centering
	\includegraphics[angle=270,width=0.45\textwidth]{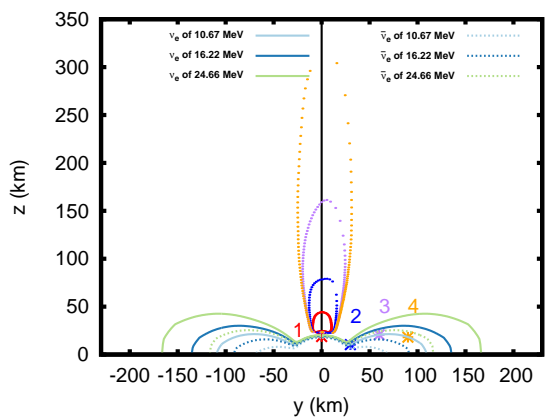}
	\caption{Matter Neutrino Resonance surfaces (dotted lines) are shown for a variety of neutrino starting locations (stars).  The colors of the surfaces are matched to the color of the star indicating the emission point, i.e. neutrinos emitted from the location indicated by the blue star encounter the blue MNR surface.  For neutrinos originating from the disk, transformation only occurs as the neutrinos pass out of the underdense funnel.  Neutrino surfaces are shown as in Fig.~\ref{fig:tra1_2}.}
	\label{mnrsurf}
\end{figure}

Following the same method, we plot MNR surfaces for neutrinos with  different starting locations on the disk in Fig.~\ref{mnrsurf}. 
For each starting point, we again use a serial of test trajectories, sweeping the plane above. We color code these surfaces corresponding to neutrino emission location. We see that neutrinos originating at the massive neutron star will all transform.  However, for the neutrinos emitted from the disk, only those that travel toward the center of the object will undergo matter neutrino resonances.  Depending on the neutrino mass hierarchy, those that do not pass over the center of the object may undergo nutation/bipolar type oscillations \cite{Dasgupta:2008cu,Malkus:2015mda}.

\section{Discussion}
\label{secDiscuss}

In the previous section  we presented the results of two flavor calculations, assuming that the relevant masses and  mixing angles are the measured values of $\delta m^2_{13}$, $\theta_{13}$ and the hierarchy is normal. In this section we consider what happens if the  hierarchy is inverted and explore the differences between two and three flavor evolution. 

\begin{figure}
	\centering
	\includegraphics[angle=270,width=0.8\textwidth]{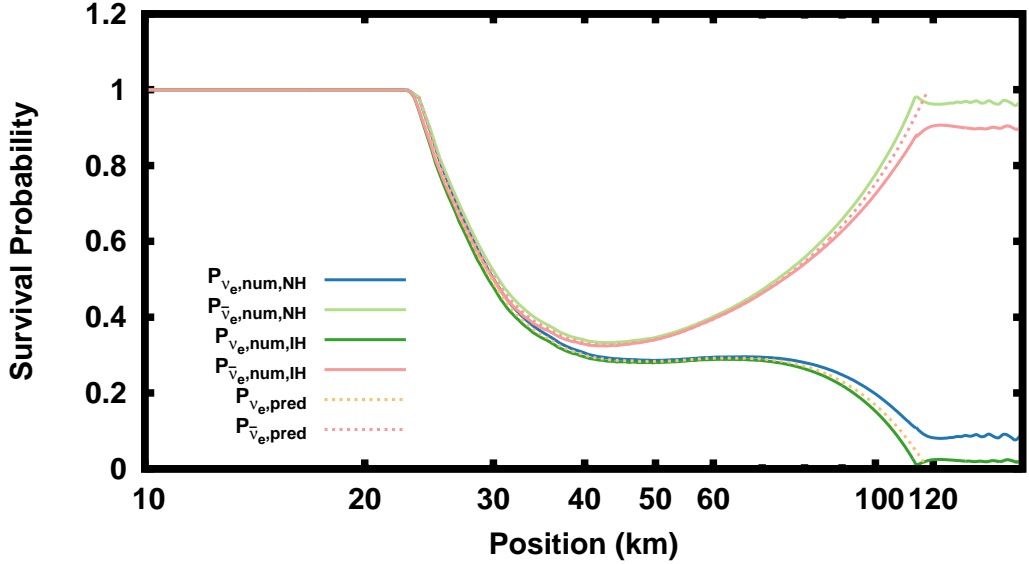}
	\caption{Flux weighted numerical survival probability (solid) and predicted survival probability (dashed) for electron neutrinos and antineutrinos along trajectory one with normal hierarchy and inverted hierarchy.}
	\label{nhih}
\end{figure}

We start with a comparison of the inverted hierarchy and the normal hierarchy.  In Ref. \cite{Vaananen:2015hfa} it was shown that the correction to the analytic prediction of the survival probabilities  from the vacuum term scales as $(\delta m^2 /E)/V_e$.  Since at the position of the matter neutrino resonance, the matter potential, $V_e$, is far above the vacuum scale,  $\delta m^2 /E$,   this correction is very small and we expect similar results in both hierarchies.  We see in  Fig.~\ref{nhih}, where we have computed the survival probability for the inverted hierarchy along trajectory one, that this expectation is confirmed by the numerics.  Both the inverted and normal hierarchies give similar results. 

\begin{figure}[ht!]
\centering
    \begin{subfigure}
            \centering
        \includegraphics[angle=270, width=3.4in]{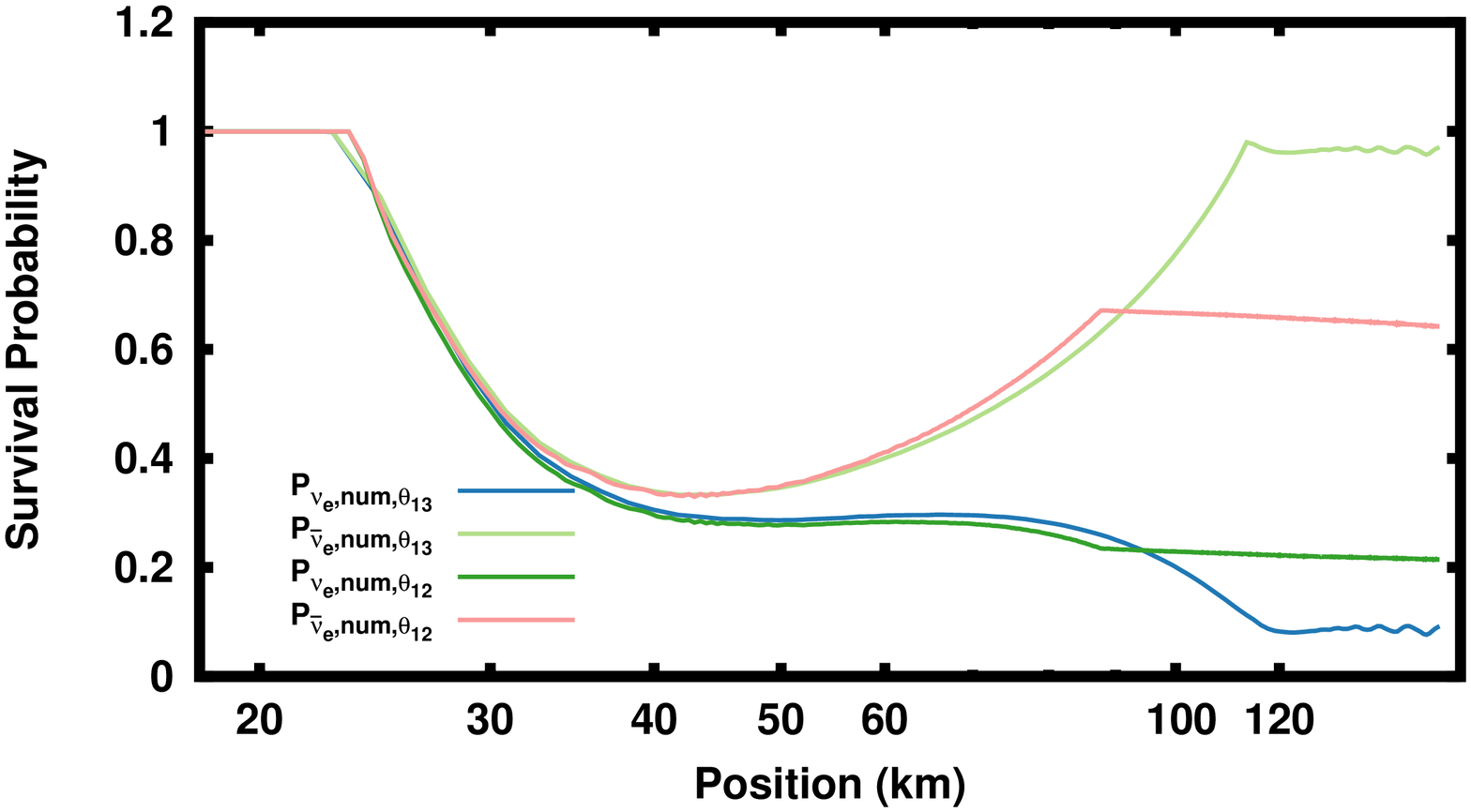}
            \end{subfigure}%
    ~ 
    \begin{subfigure}
        \centering
        \includegraphics[angle=270, width=3.4in]{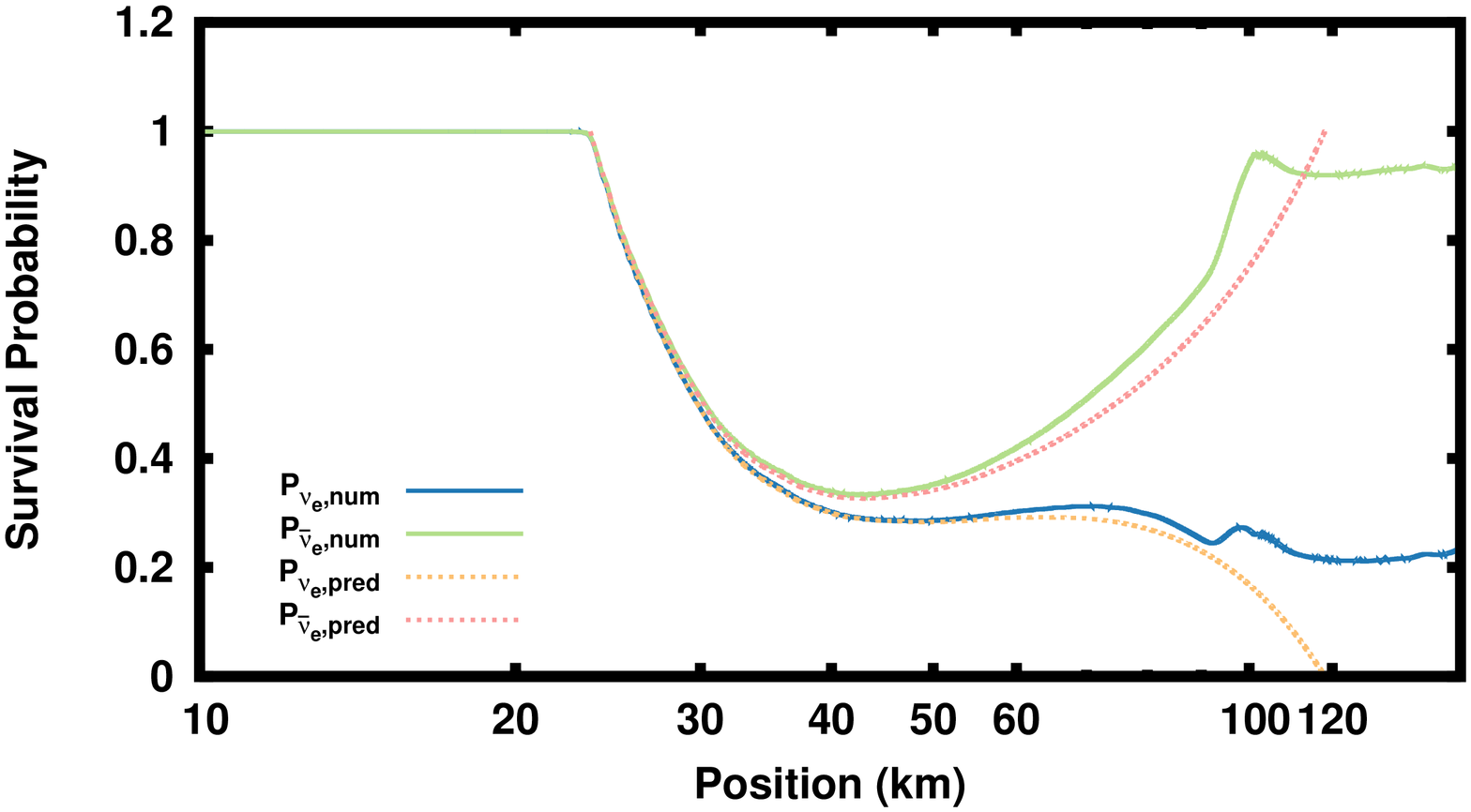}
    \end{subfigure}
        \caption{Left panel: Flux weighted numerical survival probabilities 
        with $\theta_{13}=0.15$ and $\delta m^2_{13}=2.43\times 10^{-3} \, {\rm eV}^2$  compared with $\theta_{12}=0.6$ and  $\delta m^2_{12}=7.59\times 10^{-5} {\rm eV}^2$ for trajectory one.  The calculation with smaller  $\delta m^2 \sin^2 2 \theta$ terminates earlier because the MNR transition would require an unphysically large value of the imaginary off diagonal components of the Hamiltonian.  
        Right panel: Flux weighted numerical survival probability for the three flavor calculation (solid) of neutrinos as compared with the two flavor prediction (dashed) along trajectory one.}
     \label{fig:mixing}
\end{figure}

Before we present the three flavor calculation, we first compare a two flavor calculation using  $\theta_{13} = 0.15 $ and $\delta m_{13}^2 =2.43\times 10^{-3} {\rm eV}^2$ with a two flavor calculation using $\theta_{12}=0.6$  and $\delta m^2_{12}=7.59\times 10^{-5} {\rm eV}^2$ \cite{Agashe:2014kda}.  While the analytic prediction of the evolution of a complete MNR transition has very little dependence on the vacuum scale, a small mixing angle has been previously shown to suppress transitions \cite{Vaananen:2015hfa}.
In Fig.~\ref{fig:mixing}, we see that the flux weighted survival probabilities for trajectory one in the scenario with $\theta_{12}$, $\delta m^2_{12}$ (solid green line for electron neutrino and solid pink line for electron anti-neutrino) track 
those of the scenario with $\theta_{13}$, $\delta m^2_{13}$ at beginning of the MNR transition, proceeding as they do with the smaller mixing angle. However, in the $\theta_{12}$, $\delta m^2_{12}$  scenario, the transition stops earlier at around $89$ km.  

Since the results of fluxed weighted survival probability from our multi-energy calculations matched well the single energy prediction in Eq.~\ref{eq:anal}, we further consider the system in terms of a single energy mode in order to explain the early termination of the MNR transition for the larger mixing angle and smaller mass squared difference. For single energy models,  it can be convenient to rewrite the evolution equations 
in terms of isospin using the notation of  Ref.~\cite{Malkus:2014iqa}. One of the dynamical equations of motion for isospin vectors $\mathbf{s}$ and $\bar{\mathbf{s}}$  is

\begin{equation}
\label{eq:Mixing}
\frac{d(s_{z} + \alpha \bar{s}_{z} )}{d l} = \Delta_V \sin 2 \theta (s_{y} - \alpha \bar{s}_{y}),
\end{equation}
where  $l$ is the distance along the trajectory. The $z$ components of the isospin vectors correspond to $s_{z}(\mathbf{x}) = P_{\nu_{e}} -0.5$ and $\bar{s}_{z}(\mathbf{x}) = 0.5 - \bar{P}_{\bar{\nu}_{e}}$, 
while the $y$ components of the isospin vectors correspond to
\begin{equation}
\begin{aligned}
s_{y}(\mathbf{x}) &= {\rm Im}[H_{\nu,xe}(\mathbf{x})]/|H_\nu|,\\
\bar{s}_{y}(\mathbf{x}) &= - {\rm Im}[H_{\bar{\nu},xe}(\mathbf{x})]/|H_{\bar{\nu}}|,
\end{aligned}
\end{equation}
where $|H_\nu|$ and $|H_{\bar{\nu}}|$ are the norms of the neutrino and antineutrino parts respectively of the total neutrino-neutrino self-interaction Hamiltonian, $H_{\nu \nu}$ in Eq.~\ref{eq:hnu}.
During a MNR transition, $s_{y} \approx - \alpha \bar{s}_{y}$ ~\cite{Malkus:2014iqa}.  A Standard MNR transition begins with survival probabilities  for neutrinos and antineutrinos of one, and when it completes, the neutrinos have a survival probability of zero and the antineutrino survival probability has returned to one. With these boundary conditions, according to Eq. \ref{eq:Mixing} the $y$ component of the isospin vector has an average value over the length of the transition of
\begin{equation}
 \langle s_{y}\rangle \approx \frac{-1}{2\delta  l \Delta_V \sin 2 \theta  },
 \label{eq:sycond}
\end{equation}
where $\delta l$ is the length of MNR.
For  trajectory one 
the complete MNR transition length
is approximately $95$ km. For  $\theta_{13}$, and $\delta m^2_{13}$, Eq. \ref{eq:sycond} predicts that $\langle s_{y}\rangle \approx - 0.03 $, while the full numerical calculation gives a flux weighted $\langle s_{y}\rangle = -0.032$. 
Since $|\mathbf s|$ has magnitude of 1/2, if an MNR transition would require a value of $|s_y| >1/2$, then the MNR transition will be suppressed.
For $\delta m^2_{13}$ and $\theta_{13}$ the left hand side of Eq. \ref{eq:sycond} is about an order of magnitude smaller than for $\delta m^2_{12}$ and $\theta_{12}$.  Therefore, in the  $\delta m^2_{12}$ and $\theta_{12}$ case $\langle s_y \rangle$ approaches its largest possible magnitude.  Taking into account that $|s_y|$ grows throughout the transition, in the scenario of trajectory one and $\delta m_{12}$, $\theta_{12}$ the MNR transition would be expected to stop before completion to prevent $s_y$ from taking on an unphysical value.

Turning now to three flavors, we expect that the MNR transition in trajectory one will proceed similarly to the two flavor $\delta m^2_{13}$ case until around $90$ km and then possibly diverge in region above about $90$ km . 
The right side of Fig.~\ref{fig:mixing} shows the three flavor calculation for trajectory one.
As anticipated, there is significant divergence from the two flavor $\delta m_{13}$ case beginning at around $90$ km, but before this point the  results of two and three flavors are quite similar.
Therefore, we expect that two flavor calculations are a good approximation to three flavors, as long as $|s_y|$ remains below approximately 1/2 for both $\delta m^2_{13} \sin 2 \theta_{13}$ and  $\delta m^2_{12} \sin 2 \theta_{12}$. 


\section{Conclusions}

We conduct the first studies of the flavor evolution of neutrinos above a binary neutron star merger remnant where the matter potential and the neutrino self interaction potential are computed self consistently from the same dynamical simulation \cite{Perego:2014fma}  in order to give as realistic a picture as is currently possible about matter neutrino resonance (MNR) transformation.  In this model,  which has both a massive neutron star and an accretion disk, matter neutrino resonances are a common phenomenon.  

Neutrinos typically pass through a resonance location at the edge of the low density funnel above the massive neutron star where the neutrino and matter potentials have the approximately the same magnitude.  Thus most neutrinos emitted from the massive neutron star, which begin in the funnel, have the opportunity to encounter a MNR.  For the disk neutrinos, 
outside of the funnel the matter potential is often too high for a MNR, so only those neutrinos that travel in the direction of the funnel encounter a resonance.  The exact locations of the resonances depend on the angle of travel of the neutrinos, but are mostly at $30$ km to $300$ km above the core.  In the context of the single angle approximation, most of the neutrinos that encounter a resonance as they leave the funnel region exhibit a transition. 

The type of  MNR transition varies between neutrino trajectories, with some neutrinos undergoing only a Standard MNR, some a Symmetric followed by a Standard MNR, and some having more of a hybrid appearance. In general, the Symmetric MNR transitions are caused by the neutrino self interaction potential changing from negative to positive, i.e. from becoming dominated by antineutrinos to dominated by neutrinos. Part of this behavior comes from the  spatial distribution of the relative number densities of neutrinos and antineutrinos.  The massive neutron star emits more antineutrinos than neutrinos, so over the central axis, $\bar{\nu}_e$ outnumber $\nu_e$.    In contrast, the disk emits more neutrinos than antineutrinos, so in some regions over the disk, the $\nu_e$ outnumber $\bar{\nu}_e$.  This effect drives the initially negative potential toward the positive as neutrinos exit the region above the massive neutron star.  In addition, there is a second effect which comes from the geometric, $1 - \cos \Theta$ factor in the potential which takes into account the angle of scattering on the ambient neutrinos.  This factor means that a more distended emission surface creates a larger contribution to the potential than a more compact emission surface.  Since the neutrino emitting surface is larger than the antineutrino emitting surface, this favors neutrinos at sufficient distance. Therefore the initially negative potential is pushed toward the positive as the neutrino travels away from the emitting surface. It is the combination of these two effects that creates the change in sign.

In most cases, at the end of the MNR transition(s), the electron  neutrinos have completely converted whereas the electron antineutrinos have started to convert but then returned to their original configuration. 
The evolution of the neutrinos during a MNR transition in typical circumstances is fairly insensitive to the hierarchy as well as  the mass squared difference and mixing angle. 
However, a sufficiently small $\delta m^2 \sin 2 \theta$ will suppress the MNR transition.  The value of $\delta m^2 \sin 2 \theta$ which suppresses the transition can  be reasonably well predicted using a single energy analysis of the growth of the imaginary component of the flavor basis Hamiltonian. 

Given the close proximity of some resonance locations to the neutrino emission surface, matter neutrino resonance transformation may have a number of consequences. It will alter the subsequent evolution of the neutrinos as well as the flavor composition of the neutrino signal. For example, in the absence of the MNR transition, one would expect a nutation/bipolar transition farther from the emission surfaces.  However, since the MNR transition alters the relative states of the neutrinos and antineutrinos, it also alters the prospects for this type of transition \cite{Malkus:2015mda}. The calculations presented here are performed in the context of the single angle approximation.  Many of the neutrinos from the remnant encounter the MNR resonance at the same location, which is encouraging from the point of view of possible mulit-angle effects.  However, a full multi-angle calculation would be required to know definitely what and how large these effects are.

MNR transformation may play a role in the dynamics of the remnant, the prospects for jet formation or on nucleosynthesis.  Since there are fewer $\mu$/$\tau$ type neutrinos than electron type, the MNR oscillation effectively removes some of the ability of the neutrinos and antineutrinos to convert neutrons to protons and vice versa.  Therefore, it is likely that this oscillation has an effect on any type of nucleosynthesis that is influenced by the neutrinos, for example, wind nucleosynthesis.  Future studies of MNR transitions in binary neutron star merger remnants are needed to elucidate these consequences, as well as to further probe the efficacy of the MNR transition itself.

\label{secConclude}

\begin{acknowledgments}
This work was supported in part by U.S. DOE Grants No.
DE-FG02-02ER41216 and DE-SC0004786 (YZ and GCM) and
by the Helmholtz-University Investigator grant No. VH-NG-825 (AP). 
It was enabled in part by the National Science Foundation under Grant No. PHY-1430152 (JINA Center for the Evolution of the Elements) (YZ and GCM).Some computations  
were supported by a grant from the Swiss National Supercomputing Centre (CSCS) under project ID s414 and s667 (AP). We thank the Institute for Nuclear Theory at the University of Washington for its hospitality and the Department of Energy for partial support during the completion of this work.

\end{acknowledgments}



\begin{thebibliography}{99}
 
 
\bibitem{Eichler:1989ve} 
  D.~Eichler, M.~Livio, T.~Piran and D.~N.~Schramm,
  Nature {\bf 340}, 126 (1989).
  
\bibitem{Ruffert:1998qg} 
  M.~Ruffert and H.~T.~Janka,
  Astron.\ Astrophys.\  {\bf 344}, 573 (1999)
  [astro-ph/9809280].
  
\bibitem{Rosswog:2003rv} 
  S.~Rosswog and M.~Liebend\"orfer,
  Mon.\ Not.\ Roy.\ Astron.\ Soc.\  {\bf 342}, 673 (2003)
  [astro-ph/0302301].


\bibitem{Just:2015dba} 
  O.~Just, M.~Obergaulinger, H.-T.~Janka, A.~Bauswein and N.~Schwarz,
  Astrophys.\ J.\  {\bf 816}, no. 2, L30 (2016)
  [arXiv:1510.04288 [astro-ph.HE]].
 
 
\bibitem{Wanajo:2014wha} 
  S.~Wanajo, Y.~Sekiguchi, N.~Nishimura, K.~Kiuchi, K.~Kyutoku and M.~Shibata,
  Astrophys.\ J.\  {\bf 789}, L39 (2014)
  [arXiv:1402.7317 [astro-ph.SR]].
  
\bibitem{Sekiguchi:2015dma} 
  Y.~Sekiguchi, K.~Kiuchi, K.~Kyutoku and M.~Shibata,
  Phys.\ Rev.\ D {\bf 91}, no. 6, 064059 (2015)
  
\bibitem{Goriely:2015fqa} 
  S.~Goriely, A.~Bauswein, O.~Just, E.~Pllumbi and H.~T.~Janka,
  Mon.\ Not.\ Roy.\ Astron.\ Soc.\  {\bf 452}, no. 4, 3894 (2015)
  [arXiv:1504.04377 [astro-ph.SR]].
  
\bibitem{Roberts:2016igt}
  L.~F.~Roberts {\it et al.},
  arXiv:1601.07942 [astro-ph.HE].
  
  
\bibitem{Surman:2003qt} 
  R.~Surman and G.~C.~McLaughlin,
  Astrophys.\ J.\  {\bf 603}, 611 (2004)
  [astro-ph/0308004].

  
\bibitem{Surman:2004sy} 
  R.~Surman and G.~C.~McLaughlin,
  Astrophys.\ J.\  {\bf 618}, 397 (2004)
  [astro-ph/0407206].
 
\bibitem{Surman:2005kf} 
  R.~Surman, G.~C.~McLaughlin and W.~R.~Hix,
  Astrophys.\ J.\  {\bf 643}, 1057 (2006)
  [astro-ph/0509365].
  
\bibitem{Dessart:2008zd} 
  L.~Dessart, C.~Ott, A.~Burrows, S.~Rosswog and E.~Livne,
  Astrophys.\ J.\  {\bf 690}, 1681 (2009)


\bibitem{Fernandez:2013tya} 
 R.~Fernández and B.~D.~Metzger,
  Mon.\ Not.\ Roy.\ Astron.\ Soc.\  {\bf 435}, 502 (2013)

\bibitem{Martin:2015hxa} 
  D.~Martin, A.~Perego, A.~Arcones, F.~K.~Thielemann, O.~Korobkin and S.~Rosswog,
  Astrophys.\ J.\  {\bf 813}, no. 1, 2 (2015)
   
\bibitem{Just:2014fka} 
  O.~Just, A.~Bauswein, R.~A.~Pulpillo, S.~Goriely and H.-T.~Janka,
  Mon.\ Not.\ Roy.\ Astron.\ Soc.\  {\bf 448}, 541 (2015)
  doi:10.1093/mnras/stv009
  [arXiv:1406.2687 [astro-ph.SR]].
  
  
\bibitem{Sekiguchi:2011zd} 
  Y.~Sekiguchi, K.~Kiuchi, K.~Kyutoku and M.~Shibata,
  Phys.\ Rev.\ Lett.\  {\bf 107}, 051102 (2011)
  [arXiv:1105.2125 [gr-qc]].
 
\bibitem{Foucart:2014nda} 
  F.~Foucart, M.~B.~Deaton, M.~D.~Duez, E.~O'Connor, C.~D.~Ott, R.~Haas, L.~E.~Kidder and H.~P.~Pfeiffer {\it et al.},
  Phys.\ Rev.\ D {\bf 90}, 024026 (2014)
  [arXiv:1405.1121 [astro-ph.HE]].
  
\bibitem{Palenzuela:2015dqa} 
  C.~Palenzuela, S.~L.~Liebling, D.~Neilsen, L.~Lehner, O.~L.~Caballero, E.~O'Connor and M.~Anderson,
  Phys.\ Rev.\ D {\bf 92}, no. 4, 044045 (2015)
  [arXiv:1505.01607 [gr-qc]].
  
\bibitem{Bernuzzi:2015opx} 
  S.~Bernuzzi, D.~Radice, C.~D.~Ott, L.~F.~Roberts, P.~Moesta and F.~Galeazzi,
  arXiv:1512.06397 [gr-qc].
  
\bibitem{Foucart:2015gaa} 
  F.~Foucart {\it et al.},
  Phys.\ Rev.\ D {\bf 93}, no. 4, 044019 (2016)
  [arXiv:1510.06398 [astro-ph.HE]].


  
\bibitem{Caballero:2009ww}
  O.~L.~Caballero, G.~C.~McLaughlin and R.~Surman,
  Phys.\ Rev.\ D {\bf 80} (2009) 123004

\bibitem{Caballero:2015cpa} 
  O.~L.~Caballero, T.~Zielinski, G.~C.~McLaughlin and R.~Surman,
  arXiv:1510.06011 [nucl-th].
  


  
\bibitem{Duan:2006an} 
  H.~Duan, G.~M.~Fuller, J.~Carlson and Y.~Z.~Qian,
  Phys.\ Rev.\ D {\bf 74}, 105014 (2006)
 
\bibitem{Hannestad:2006nj} 
  S.~Hannestad, G.~G.~Raffelt, G.~Sigl and Y.~Y.~Y.~Wong,
  Phys.\ Rev.\ D {\bf 74}, 105010 (2006)
  Erratum: [Phys.\ Rev.\ D {\bf 76}, 029901 (2007)]
  [astro-ph/0608695].

  
\bibitem{Balantekin:2006tg} 
  A.~B.~Balantekin and Y.~Pehlivan,
  J.\ Phys.\ G {\bf 34}, 47 (2007)
   
  
\bibitem{EstebanPretel:2008ni} 
  A.~Esteban-Pretel, A.~Mirizzi, S.~Pastor, R.~Tomas, G.~G.~Raffelt, P.~D.~Serpico and G.~Sigl,
  Phys.\ Rev.\ D {\bf 78}, 085012 (2008)
  
\bibitem{Gava:2009pj} 
  J.~Gava, J.~Kneller, C.~Volpe and G.~C.~McLaughlin,
  Phys.\ Rev.\ Lett.\  {\bf 103}, 071101 (2009)
  [arXiv:0902.0317 [hep-ph]].
  
\bibitem{Duan:2010bg} 
  H.~Duan, G.~M.~Fuller and Y.~Z.~Qian,
  Ann.\ Rev.\ Nucl.\ Part.\ Sci.\  {\bf 60}, 569 (2010)
  [arXiv:1001.2799 [hep-ph]].
  
\bibitem{Duan:2010af} 
  H.~Duan, A.~Friedland, G.~C.~McLaughlin and R.~Surman,
  J.\ Phys.\ G {\bf 38}, 035201 (2011)

\bibitem{Pehlivan:2011hp} 
  Y.~Pehlivan, A.~B.~Balantekin, T.~Kajino and T.~Yoshida,
  Phys.\ Rev.\ D {\bf 84}, 065008 (2011)


\bibitem{Cherry:2012zw} 
  J.~F.~Cherry, J.~Carlson, A.~Friedland, G.~M.~Fuller and A.~Vlasenko,
  Phys.\ Rev.\ Lett.\  {\bf 108}, 261104 (2012)
  
\bibitem{Volpe:2013jgr} 
  C.~Volpe, D.~V\"a\"an\"anen and C.~Espinoza,
  Phys.\ Rev.\ D {\bf 87}, no. 11, 113010 (2013)
  
\bibitem{Vlasenko:2014bva} 
  A.~Vlasenko, G.~M.~Fuller and V.~Cirigliano,
  arXiv:1406.6724 [astro-ph.HE].

  
  
\bibitem{Dasgupta:2008cu} 
  B.~Dasgupta, A.~Dighe, A.~Mirizzi and G.~G.~Raffelt,
  Phys.\ Rev.\ D {\bf 78}, 033014 (2008)
  
\bibitem{Malkus:2015mda} 
  A.~Malkus, G.~C.~McLaughlin and R.~Surman,
  Phys.\ Rev.\ D {\bf 93}, no. 4, 045021 (2016)
  [arXiv:1507.00946 [hep-ph]].
  
  
\bibitem{Malkus:2012ts} 
  A.~Malkus, J.~P.~Kneller, G.~C.~McLaughlin and R.~Surman,
  Phys.\ Rev.\ D {\bf 86}, 085015 (2012)
  [arXiv:1207.6648 [hep-ph]].
 
\bibitem{Malkus:2014iqa} 
  A.~Malkus, A.~Friedland and G.~C.~McLaughlin,
  arXiv:1403.5797 [hep-ph].
  
\bibitem{Vaananen:2015hfa} 
  D.~Vaananen and G.~C.~McLaughlin,
  Phys.\ Rev.\ D {\bf 93}, no. 10, 105044 (2016)
  [arXiv:1510.00751 [hep-ph]].
  
  
\bibitem{Wu:2015fga} 
  M.~R.~Wu, H.~Duan and Y.~Z.~Qian,
  Phys.\ Lett.\ B {\bf 752}, 89 (2016)
  

  
\bibitem{Stapleford:2016jgz} 
  C.~J.~Stapleford, D.~J.~Väänänen, J.~P.~Kneller, G.~C.~McLaughlin and B.~T.~Shapiro,
  arXiv:1605.04903 [hep-ph].
  
\bibitem{Perego:2014fma} 
  A.~Perego, S.~Rosswog, R.~M.~Cabez\'on, O.~Korobkin, R.~K\"appeli, A.~Arcones and M.~Liebend\"orfer,
  Mon.\ Not.\ Roy.\ Astron.\ Soc.\  {\bf 443}, no. 4, 3134 (2014)
  [arXiv:1405.6730 [astro-ph.HE]].
 
\bibitem{Perego:2015agy}
  A.~Perego, R.~Cabez\'on and R.~K\"appeli,
  Astrophys.\ J.\ Suppl.\  {\bf 223} (2016) no.2,  22
  [arXiv:1511.08519 [astro-ph.IM]].
  
\bibitem{Kneller:2009vd} 
  J.~P.~Kneller and G.~C.~McLaughlin,
  Phys.\ Rev.\ D {\bf 80}, 053002 (2009)
  [arXiv:0904.3823 [hep-ph]].
  
\bibitem{Agashe:2014kda} 
  K.~A.~Olive {\it et al.} [Particle Data Group Collaboration],
  Chin.\ Phys.\ C {\bf 38}, 090001 (2014).

\bibitem{An:2013zwz} 
  F.~P.~An {\it et al.} [Daya Bay Collaboration],
  Phys.\ Rev.\ Lett.\  {\bf 112}, 061801 (2014)
  [arXiv:1310.6732 [hep-ex]].
  
  
\end{thebibliography}
\end{document}